\PassOptionsToPackage{unicode}{hyperref}
\PassOptionsToPackage{hyphens}{url}
\PassOptionsToPackage{dvipsnames,svgnames,x11names}{xcolor}
\documentclass[
  12pt]{article}

\usepackage{graphicx}
\usepackage{hyphenat}
\usepackage{xcolor}         
\usepackage{amssymb}
\usepackage{natbib,graphicx,setspace,lscape,longtable}
\usepackage{epsfig,graphicx}
\usepackage{extarrows}
\usepackage{lscape}
\usepackage{subfigure}
\usepackage{subcaption}
\usepackage{amsmath,amsthm,amssymb,amsfonts,color}
\usepackage{booktabs}
\usepackage{tabularx}
\usepackage{epstopdf}
\usepackage{amsthm}
\usepackage{bm}
\usepackage{threeparttable}
\usepackage{multirow}
\usepackage{enumerate}
\usepackage{enumitem} 
\usepackage{algorithm,setspace}
\usepackage{algorithmic}
\usepackage{placeins}

\newtheorem{theorem}{Theorem}
\newtheorem{lemma}{Lemma}
\newtheorem{proposition}{Proposition}

\newtheorem{corollary}{Corollary}
\newtheorem{definition}{Definition}

\def\beq{\begin{equation}}
\def\eeq{\end{equation}}
\def\beqr{\begin{eqnarray}}
\def\eeqr{\end{eqnarray}}
\def\beqrs{\begin{eqnarray*}}
\def\eeqrs{\end{eqnarray*}}
\def\bet{\begin{theorem}}
\def\eet{\end{theorem}}
\def\bel{\begin{lemma}}
\def\eel{\end{lemma}}
\def\bed{\begin{definition}}
\def\eed{\end{definition}}
\def\bep{\begin{proposition}}
\def\eep{\end{proposition}}
\def\bg{\begin{figure}[tbph]\begin{center}}
\def\eg{\end{center}\end{figure}}

\def\bc{\begin{center}}
\def\ec{\end{center}}

\def\wt{\widetilde}

\def\wh{\widehat}

\def\mN{\mathcal{N}}

\def\mR{\mathbb{R}}
\def\mI{\mathcal I}
\def\mS{\mathbb S}

\def\mS{\mathcal S}

\def\mT{\mathcal T}

\def\mS{\mathcal S}

\def\BIC{\mbox{BIC}}

\def\beq{\begin{equation}}
\def\eeq{\end{equation}}
\def\beqr{\begin{eqnarray}}
\def\eeqr{\end{eqnarray}}
\def\beqrs{\begin{eqnarray*}}
\def\eeqrs{\end{eqnarray*}}

\usepackage{amsmath,amssymb}

\usepackage[T1]{fontenc}
\usepackage[utf8]{inputenc}

\usepackage{lmodern}
\ifPDFTeX\else  
\fi
\IfFileExists{upquote.sty}{\usepackage{upquote}}{}
\IfFileExists{microtype.sty}{
  \usepackage[]{microtype}
  \UseMicrotypeSet[protrusion]{basicmath} 
}{}
\makeatletter
\@ifundefined{KOMAClassName}{
  \IfFileExists{parskip.sty}{%
    \usepackage{parskip}
  }{
    \setlength{\parindent}{0pt}
    \setlength{\parskip}{6pt plus 2pt minus 1pt}}
}{
  \KOMAoptions{parskip=half}}
\makeatother
\usepackage{xcolor}
\makeatletter
\ifx\paragraph\undefined\else
  \let\oldparagraph\paragraph
  \renewcommand{\paragraph}{
    \@ifstar
      \xxxParagraphStar
      \xxxParagraphNoStar
  }
  \newcommand{\xxxParagraphStar}[1]{\oldparagraph*{#1}\mbox{}}
  \newcommand{\xxxParagraphNoStar}[1]{\oldparagraph{#1}\mbox{}}
\fi
\ifx\subparagraph\undefined\else
  \let\oldsubparagraph\subparagraph
  \renewcommand{\subparagraph}{
    \@ifstar
      \xxxSubParagraphStar
      \xxxSubParagraphNoStar
  }
  \newcommand{\xxxSubParagraphStar}[1]{\oldsubparagraph*{#1}\mbox{}}
  \newcommand{\xxxSubParagraphNoStar}[1]{\oldsubparagraph{#1}\mbox{}}
\fi
\makeatother

\usepackage{longtable,booktabs,array}
\usepackage{calc} 
\usepackage{etoolbox}
\makeatletter
\patchcmd\longtable{\par}{\if@noskipsec\mbox{}\fi\par}{}{}
\makeatother
\IfFileExists{footnotehyper.sty}{\usepackage{footnotehyper}}{\usepackage{footnote}}
\makesavenoteenv{longtable}
\usepackage{graphicx}
\makeatletter
\def\maxwidth{\ifdim\Gin@nat@width>\linewidth\linewidth\else\Gin@nat@width\fi}
\def\maxheight{\ifdim\Gin@nat@height>\textheight\textheight\else\Gin@nat@height\fi}
\makeatother
\setkeys{Gin}{width=\maxwidth,height=\maxheight,keepaspectratio}
\makeatletter
\def\fps@figure{htbp}
\makeatother

\makeatletter
\@ifpackageloaded{caption}{}{\usepackage{caption}}
\AtBeginDocument{%
\ifdefined\contentsname
  \renewcommand*\contentsname{Table of contents}
\else
  \newcommand\contentsname{Table of contents}
\fi
\ifdefined\listfigurename
  \renewcommand*\listfigurename{List of Figures}
\else
  \newcommand\listfigurename{List of Figures}
\fi
\ifdefined\listtablename
  \renewcommand*\listtablename{List of Tables}
\else
  \newcommand\listtablename{List of Tables}
\fi
\ifdefined\figurename
  \renewcommand*\figurename{Figure}
\else
  \newcommand\figurename{Figure}
\fi
\ifdefined\tablename
  \renewcommand*\tablename{Table}
\else
  \newcommand\tablename{Table}
\fi
}
\@ifpackageloaded{float}{}{\usepackage{float}}
\floatstyle{ruled}
\@ifundefined{c@chapter}{\newfloat{codelisting}{h}{lop}}{\newfloat{codelisting}{h}{lop}[chapter]}
\floatname{codelisting}{Listing}

\makeatother
\makeatletter
\@ifpackageloaded{caption}{}{\usepackage{caption}}
\@ifpackageloaded{subcaption}{}{\usepackage{subcaption}}
\makeatother

\ifLuaTeX
  \usepackage{selnolig}  
\fi
\usepackage[]{natbib}
\usepackage{bookmark}

\IfFileExists{xurl.sty}{\usepackage{xurl}}{} 
\hypersetup{
  pdftitle={Title},
  pdfauthor={Author 1; Author 2},
  pdfkeywords={3 to 6 keywords, that do not appear in the title},
  colorlinks=true,
  linkcolor={blue},
  filecolor={Maroon},
  citecolor={Blue},
  urlcolor={Blue},
  pdfcreator={LaTeX via pandoc}}

\newcommand{\anon}{1}

\begin{document}

\def\spacingset#1{\renewcommand{\baselinestretch}%
{#1}\small\normalsize} \spacingset{0.95}


\if1\anon
{
  \title{\bf Common-Individual Embedding for Dynamic Networks with Temporal Group Structure}
  \author{
  Hairi Bai \thanks{
    The authors are listed in the alphabetic order. }
    \\
    School of Economics, Xiamen University\\
    Xinyan Fan \thanks{Xinyan Fan's research was supported by the National Natural Science Foundation of China (12201626, 72571272).} \\
    School of Statistics,
    Renmin University of China\\
  Kuangnan Fang \thanks{Kuangnan Fang's research was supported by the National Natural Science Foundation of China (12571313).} \\
    School of Economics, Xiamen University\\
    and \\
    Yan Zhang \footnotemark[1] \\
    School of Statistics and Information, \\
    Shanghai University of International Business and Economics}
  \maketitle
} \fi

\if0\anon
{
  \bigskip
  \bigskip
  \bigskip
  \begin{center}
    {\LARGE\bf Common-Individual Embedding for Dynamic Networks with Temporal Group Structure}
\end{center}
  \medskip
} \fi

\bigskip
\begin{abstract}
We propose STANE (Shared and Time-specific Adaptive Network Embedding), a new joint embedding framework for dynamic networks that captures both stable global structures and localized temporal variations. To further improve the model’s adaptability to transient changes, we introduce Sparse STANE, which models time-specific changes as sparse perturbations, thereby improving interpretability. Unlike existing methods that either overlook cross-time similarities or enforce overly smooth evolution, Sparse STANE integrates temporal clustering with sparse deviation modeling to strike a flexible balance between persistence and change. We also provide non-asymptotic error guarantees of embeddings and show that our estimator can reliably identify changed node pairs when deviations are sparse. On synthetic and real-world political conflict networks, STANE and its extensions improve temporal clustering accuracy and structural recovery, outperforming state-of-the-art baselines. These results highlight the potential of STANE in applications such as international relations modeling, where both persistent and transient connections matter. Our findings underscore the power of structured dynamic embeddings for revealing interpretable patterns in network evolution.
\end{abstract}

\noindent%
{\it Keywords:}  Network embedding,  Temporal clustering, Sparse perturbations
\vfill

\newpage
\spacingset{1.8} 

\section{Introduction}
\label{intro}
Networks provide a powerful framework for modeling complex relationships and interactions in various domains, such as social connections \citep{de2016learning, chen2024monitoring}, financial transactions \citep{giudici2016graphical, pesaran2020econometric}, and biological systems \citep{girvan2002community, kolaczyk2014statistical}. Traditional network analysis primarily focuses on static structures, assuming that connections remain unchanged over time. However, many real-world systems are inherently dynamic, where relationships evolve continuously. A dynamic network is typically represented as a sequence of network snapshots over discrete time steps, where different networks share a common node set and edges may appear, disappear, or change in strength \citep{krivitsky2014separable, sewell2015latent, xue2022dynamic}. Dynamic networks introduce new challenges and opportunities, as they not only capture structural dependencies but also temporal patterns that govern how interactions change over time. Understanding these temporal dynamics is crucial for key applications such as forecasting future connections \citep{chen2019lstm, chen2021time}, detecting structural shifts \citep{cheung2020simultaneous, padilla2022change}, and identifying anomalous behaviors \citep{10.1145/3219819.3220024}.

\begin{figure}[t]
    \centering
    \includegraphics[width=\textwidth]{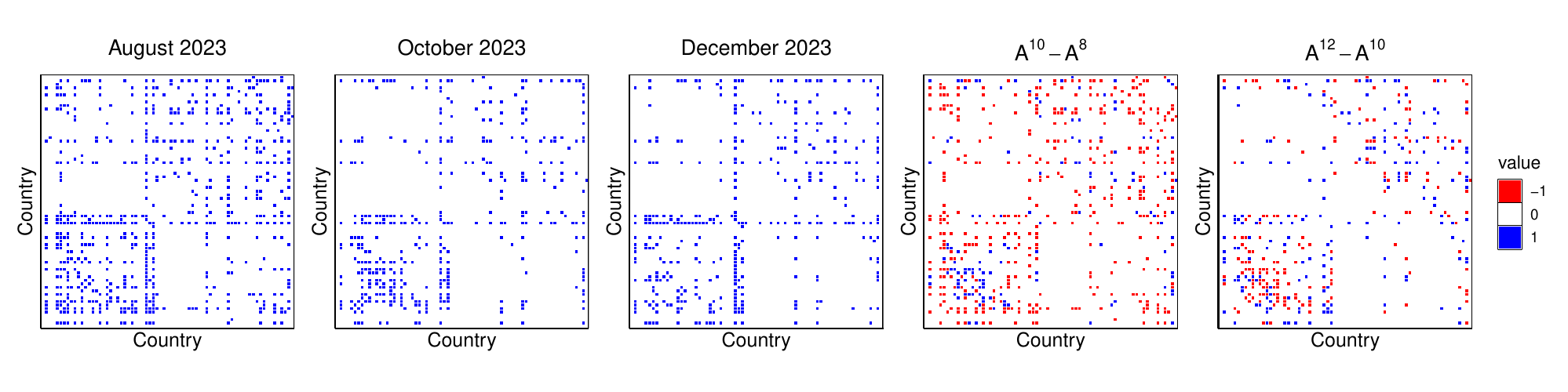}
    \caption{Adjacency matrices of political networks at three time points (Panels 1–3), and corresponding difference matrices (Panels 4–5).}
    \label{fig:intro}
\end{figure}

For dynamic networks, some approaches either treat each timestamp independently, failing to leverage shared characteristics across time periods \citep{10.1145/3184558.3191526, zhang2024change}, or arbitrarily merge networks over certain time spans, thereby limiting adaptability to sudden changes \citep{bhattacharjee2020change, xu2022statistical}. This motivates the need to develop a more flexible model that balances two important aspects. One is capturing stable, global structures that persist over time, while the other is effectively modeling time-specific variations. Achieving this balance is especially crucial for network data encompassing multiple regions worldwide, such as political conflict networks or trade flow networks at the global level. \citep{chen2019lstm,Cepeda-López19052019}. As a motivating example, consider a political network in which each node corresponds to a region, while an edge between two regions indicates the existence of  a conflict, either verbal or material, within a certain time period. Figure \ref{fig:intro} displays three binary structural political networks along with the differences in their adjacency matrices. The first three panels show the heatmap of the adjacency matrices for August 2023, October 2023 and December 2023, where purple squares indicate the presence of an edge between two regions. The last two panels illustrate the differences between these matrices. It is obvious that these three time periods share a common underlying structure, while also exhibiting time-specific connections, indicating the existence of the stable and time-varying structure across different time periods. In this context, it is particularly interesting to study the stable and evolving relationships between countries in a network over time. Shared connections provide important insights into evolutionarily conserved structures in political relations. For example, some countries have maintained stable conflicts over time. However, international relations are constantly evolving. The time-specific structure of political networks can help better understand changes and infer their causes. The focus of this paper is to develop a new model to capture both the shared and unique structures of dynamic networks.

Recent studies about dynamic networks have also expanded on these challenges by incorporating both time-invariant
and time-varying structures. For example, \cite{zhang2024change} propose a tensor-based embedding model that decomposes the probability tensor into time-invariant latent node embeddings and time-specific variations.  Although our primary focus is on dynamic networks, they can be viewed as a special case of multilayer networks, with each time step corresponding to a layer. This connection has inspired methods for
disentangling shared and layer-specific structures. For example,  \cite{wang2019common} decompose the logit of adjacency matrices into a shared baseline and a low-rank layer-specific deviation. Similarly
\cite{arroyo2021inference} propose the COSIE model, which captures network-specific variations while assuming a shared latent subspace. Building on these ideas, \cite{Agterberg28072025} improve clustering accuracy by leveraging shared community structures across time. \cite{gollini2016joint} and \cite{d2019latent} propose latent space models with shared node embeddings and layer-specific parameters to capture global properties like density and homophily. However, such shared embeddings assume static node behavior, which is often unrealistic. To increase flexibility, \cite{pmlr-v119-zhang20aa} introduce layer-specific degree heterogeneity and connection matrices, though latent positions remain fixed. Further,  \cite{macdonald2022latent} develope the MultiNeSS model to better separate common and individual structures, while \cite{tian2024efficient} consider the unified framework for latent space model of heterogeneous network through decomposing adjacency matrices into shared and network-specific latent components. However, these methods ignore temporal continuity and grouped patterns of change. In real-world networks, structures often evolve gradually, with stable phases and abrupt transitions (i.e., change points) triggered by external events. For instance, as shown in Figure \ref{intro}, the connection patterns of the political conflict networks differ across three observed periods, with the change from August to October 2023 being larger than that from October to December 2023. This indicates temporal dynamics across periods and suggests the presence of group structures that persist over time. Moreover, the differences among these three periods are sparse, highlighting the sparsity of the time-varying components.

In this work, we propose a joint network embedding framework, {\it STANE (Shared and Time-specific Adaptive Network Embedding)}, along with its sparse extension, {\it Sparse STANE}, to capture both shared and time-varying structures in dynamic networks. Our main contributions are:

\begin{itemize}
    \item Address two key limitations of existing dynamic network models by introducing a framework that (i) incorporates a temporal clustering mechanism to detect latent group structures across time, and (ii) explicitly models the sparsity of time-varying components, allowing only a small subset of node pairs to change over time. These innovations enhance the model’s flexibility in capturing both global and localized dynamics.

    \item Develop an efficient estimation algorithm based on projected gradient descent, which demonstrates stable performance. To further improve robustness, we design an initialization procedure that yields high-quality starting values.

    \item Establish non-asymptotic error bounds for the proposed estimator, characterizing its finite-sample behavior under mild conditions. Additionally, prove that when changes are sparse, the set of changed nodes can be identified with high probability, showcasing the model’s ability to detect localized temporal perturbations.

    \item Validate the proposed method through extensive experiments on both synthetic and real-world political conflict networks. Results show that STANE and its extensions effectively recover temporal clusters, identify structural shifts, and offer interpretable insights into evolving international relations. Comparisons with state-of-the-art methods further confirm the superior performance of STANE in structural recovery.
\end{itemize}

The rest of this article is organized as follows. Section 2 introduces the STANE model along with its sparse extension and the corresponding estimation procedures. Section 3 presents the theoretical properties of the estimators. Section 4 and Section 5 demonstrate the efficiency of the proposed STANE framework through extensive simulation studies and two real world examples. At last, we conclude the article in Section 6.

\section{Models and methodology}
\label{Method}

\subsection{Notations}
\label{notations}

We adopt the following general notations throughout this paper. For any vector \( v = (v_1, \dots, v_p)^\top \in \mathbb{R}^p \), let \( v_i \) denote its \( i \)th element, and \( \|v\|_2 \) denote the $\ell_2$ norm of \( v \).  
For any matrix \( M \in \mathbb{R}^{n_1 \times n_2} \), we use \( M_{ij} \), \( M_{i\cdot} \), and \( M_{\cdot j} \) to denote the \((i,j)\)th entry, the \( i \)th row, and the \( j \)th column of \( M \), respectively. Let \( \sigma_i(M) \) denote the \( i \)th largest singular value of \( M \), and \( \lambda_i(M) \) denote its \( i \)th largest eigenvalue in magnitude. We use \( \|M\|_{\max} \), \( \|M\|_{\min} \) to denote the maximum and minimum absolute entry of \( M \), and \( \|M\|_F \), \( \|M\|_2 \) to denote the Frobenius norm and spectral norm of \( M \), respectively.  For any 3-way tensor \( \bm{T} = [T^{(1)}; \dots; T^{(n_3)}] \in \mathbb{R}^{n_1 \times n_2 \times n_3} \), let \( T^{(k)} \) denote the \( k \)th frontal slice of \( \bm{T} \), for \( k = 1, \dots, n_3 \). The Frobenius norm of \( \bm{T} \) is defined as $\|\bm{T}\|_F = \left\{ \sum_{i=1}^{n_1} \sum_{j=1}^{n_2} \sum_{k=1}^{n_3} \left( T_{ij}^{(k)} \right)^2 \right\}^{1/2}$. For any two tensors \( \bm{A}, \bm{B} \in \mathbb{R}^{n_1 \times n_2 \times n_3} \), their inner product is defined as
$\langle \bm{A}, \bm{B} \rangle = \sum_{i=1}^{n_1} \sum_{j=1}^{n_2} \sum_{k=1}^{n_3} A_{ij}^{(k)} B_{ij}^{(k)}$. We use symbol \( \circ\) to denote the outer product of vectors. That is, for any 3 vectors \(x\in\mR^{n_1}\), \(y\in\mR^{n_2}\), \(z\in\mR^{n_3}\), $\bm{T}=x\circ y\circ z\in \mR^{n_1\times n_2 \times n_3}$ and $\bm{T}^{(k)}_{ij}=x_iy_jz_k$. For any real-valued function \( f(\cdot) \), we define \( f(M) \) and \( f(\bm{T}) \) as the shorthand for applying \(f(\cdot)\) element-wisely to  \( M \) and \( \bm{T} \), respectively. That is $f(M)\in\mR^{n_1\times n_2}$ and $f(\bm{T})\in\mR^{n_1\times n_2\times n_3}$ with entries given by \( f(M)_{ij} = f(M_{ij}) \) and \( f(\bm{T})^{(k)}_{ij} = f(\bm{T}^{(k)}_{ij}) \).
In addition, let \( I_p \) denote the \( p \times p \) identity matrix; \(\mI(\cdot)\) denote the indicator function; \( \bm{1}_p \), \( \bm{0}_p \) denote the \( p \)-dimensional all-ones and all-zeros vectors, respectively; and \( \bm{0}_{n_1 \times n_2} \) denote the \( n_1 \times n_2 \) zero matrix.

\subsection{Model}
\label{model}

Suppose we observe a dynamic, undirected network over \( T \) time points, defined on a common set of \( N \) nodes indexed by \( \mathcal{N} = \{1, \dots, N\} \). At each time point \( t \in \{1, \dots, T\} \), the network structure is represented by an adjacency matrix \( A^{(t)} = (A_{ij}^{(t)}) \in \{0,1\}^{N \times N} \), where \( A_{ij}^{(t)} = A_{ji}^{(t)} = 1 \) indicates the presence of an edge between nodes \( i \) and \( j \) at time \( t \), and \( A_{ij}^{(t)} = A_{ji}^{(t)} = 0 \) otherwise. We assume there are no self-loops, i.e., \( A_{ii}^{(t)} = 0 \) for all \( i \) and \( t \). By stacking the \( T \) adjacency matrices, we obtain a three-way adjacency tensor \( \bm{A} = [A^{(1)}; \dots; A^{(T)}] \in \{0,1\}^{N \times N \times T} \), where each entry \( A_{ij}^{(t)} \) is conditionally independent and sampled from a Bernoulli distribution with probability \( P_{ij}^{(t)} \). Let $P^{(t)} = (P_{ij}^{(t)})$ and \( \bm{P} = [P^{(1)}; \dots; P^{(T)}] \in \mathbb{R}^{N \times N \times T} \) denote the underlying probability tensor.

In many real-world dynamic networks, the underlying structure often exhibits both stable and fluctuating components. For example, long-term relationships such as persistent collaborations or stable social ties may remain largely unchanged over time, while short-term events or external shocks can lead to sudden changes in connectivity patterns. Motivated by this, we assume that the dynamic network structure can be decomposed into two components: a time-invariant common structure \( \bm{S} \), and a time-varying individual structure \( \bm{D} \), such that
\begin{align}
    \text{logit}(\bm{P}) = \bm{\Theta} = \bm{S} + \bm{D}.
    \label{model1}
\end{align}
We model \( \bm{S} \) and \( \bm{D} \) via CP decomposition:
\[
\bm{S} = \sum_{r=1}^{R_S} z_r \circ z_r \circ \bm{1}_T, \quad
\bm{D} = \sum_{k=1}^{K} \sum_{r=1}^{R_D} u_r^{(k)} \circ u_r^{(k)} \circ v_r^{(k)},
\]

Here, each slice of \( \bm{S} = [S^{(1)}; \dots; S^{(T)}] \in \mathbb{R}^{N \times N \times T} \) is identical and equal to \( S = Z Z^\top \), where \( Z = (z_1,\dots,z_{R_S}) \in \mathbb{R}^{N \times R_S} \). The \( i \)th row of \( Z \), denoted \( Z_i \), represents the time-invariant latent position of node \( i \). The inner product \( Z_i^\top Z_j \) reflects the proximity of nodes \( i \) and \( j \): a larger inner product indicates a higher probability of connection. This structure is known as the projection model, originally proposed by \cite{hoff2002latent}.

To capture heterogeneity across time, we assume that the \( T \) time points can be grouped into \( K \) temporal groups. Let \( \mathcal{G} = \{g_1,\dots,g_T\} \in \{1,\dots,K\}^T \) denote the group labels, where \( g_t = k \) indicates that time point \( t \) belongs to group \( k \).
The dynamic part \( \bm{D} = [D^{(1)};\dots;D^{(T)}] \) models time-varying variations. For each group \( k \), let \( u_r^{(k)} \in \mathbb{R}^N \) capture node-level variations, and \( v_r^{(k)} = (v_{r1}^{(k)},\dots,v_{rT}^{(k)})^\top \in \mathbb{R}^T \) describe the temporal influence pattern. Define \( U^{(k)} = (u_1^{(k)}, \dots, u_{R_D}^{(k)}) \in \mathbb{R}^{N \times R_D} \); its \( i \)th row captures the latent position of node \( i \) within time group \( k \). The vector \( v_r^{(k)} \) determines how the latent variation affects the probability at different time points. To reflect group structure, we assume \( v_{rt}^{(k)} = 0 \) when \( g_t \neq k \).

To further simplify Model \eqref{model1} into a matrix form, we define \( V^{(t)} \in \mathbb{R}^{R_D \times R_D} \) as a diagonal matrix, where the diagonal elements are given by \( \{v^{(k)}_{1t}, \dots, v^{(k)}_{R_D t}\} \) for time point \( t \) with \( g_t = k \). Then, the model at time \( t \) can be written as:
\begin{gather}
    \operatorname{logit}(P^{(t)}) = \Theta^{(t)} = S + D^{(t)}, \nonumber \\
    \text{with}~S=ZZ^\top,~~ D^{(t)}=U^{(g_t)}V^{(t)}U^{(g_t)\top}.
    \label{model2}
\end{gather}

We refer to Model \eqref{model1} as the Shared and Time-specific Adaptive Network Embedding (STANE) model. Note that STANE is not identifiable with respect to the parameters \( Z \), \( U^{(k)} \), and \( V^{(t)} \) without additional constraints. The following proposition establishes the conditions under which the model becomes identifiable. The proof is provided in the Appendix A.1.

\bep 
\label{pro1}
Suppose that two sets of parameters $(Z; ~\{U^{(k)}\}_{k=1}^{K};~ \{V^{(t)}\}_{t=1}^{T};~\mathcal{G})$ and $(Z_\dagger; ~\{U^{(k)}_\dagger\}_{k=1}^{K};~ \{V^{(t)}_\dagger\}_{t=1}^{T};~\mathcal{G}_\dagger)$ satisfy the following conditions:
\begin{itemize}
    \item Define an undirected graph \(\mathcal{G}_I\) over the network layers, where each vertex corresponds to a time index \(t = 1, \dots, T\). An edge is placed between time points \(t_1\) and \(t_2\) if the matrix $(Z,\ U^{(g_{t_1})} V^{(t_1)1/2},\  U^{(g_{t_2})} V^{(t_2)1/2}) \in \mR^{N \times (R_S + 2R_D)}$ has full column rank, i.e., its columns are linearly independent. We assume that the graph \(\mathcal{G}_I\) is connected.

    \item $U^{(k)\top}U^{(k)}/N=I_{R_D}$ for $k=1,...,K$.
    \item For every $k$, at least one of the diagonal matrix $V^{(t)}$ with $g_t=k$ is full rank. 
    
\end{itemize}
Then, given $g_t=k$ and
$ZZ^\top+U^{(k)}V^{(t)}U^{(k)\top}=Z_\dagger Z_\dagger^\top+U^{(k)}_\dagger V^{(t)}_\dagger U^{(k)\top}_\dagger$, we can conclude that there exist orthogonal matrices $O\in\mR^{R_S\times R_S}$ and  $O^{(k)}\in \mR^{R_D\times R_D}$ for $k=1,...,K$, such that 
\beq
Z=Z_\dagger O, ~ U^{(k)}=U_\dagger^{(k)}O^{(k)},~ \text{and}~V^{(t)}=O^{(k)\top} V^{(t)}_\dagger O^{(k)}. \notag
\eeq
\eep

\subsubsection{Model Extension: Sparse STANE }

In many empirical scenarios, while some interactions among nodes evolve over time, others remain relatively stable throughout some observation periods. For example, in dynamic social networks, long-standing relationships may persist with little variation, whereas transient or newly-formed ties exhibit more temporal dynamics. Motivated by such patterns, we extend the STANE model to allow for partially time-invariant interactions.

Specifically, we consider the case where some blocks of the time-varying component \( D^{(t)} = U^{(g_t)} V^{(t)} U^{(g_t)\top} \) are identically zero, indicating that certain nodes do not exhibit any temporal variation in their interaction patterns. Formally, we assume that for some nodes \( i \in \{1, \dots, N\} \), the corresponding row in the group-specific matrix \( U^{(k)} \) is zero, i.e.,
\[
U^{(k)\top}_{i\cdot} = \bm{0}_{R_D},
\]
which implies that the \( i \)th row and column of \( D^{(t)} \) are zero for all \( t \) such that \( g_t = k \). The key difference from the STANE model is that we now allow for row-wise zero patterns in \( U^{(k)} \), reflecting nodes whose latent positions do not vary over time.

This model, referred to as the Sparse STANE, provides a more flexible framework to capture heterogeneous dynamics across nodes and time, while still maintaining the interpretability and structural clarity of the original STANE formulation.

\subsection{Parameter estimation}
\label{parameter estimation}
\subsubsection{Projected gradient descent method}
\label{pgd}
In this section, we first present the parameter estimation procedure for the proposed STANE model. Denote $\mT=(Z; \{U^{(k)}\}_{k=1}^{K}; \{V^{(t)}\}_{t=1}^T;\mathcal{G})$, then the negative log-likelihood function for the observed networks is given by:
\beq
    \ell_1(\mT)=-\sum_{t=1}^{T}\sum_{i=1}^{N}\sum_{j=1}^{N} \left[A_{ij}^{(t)}\Theta_{ij}^{(t)}-\log\left\{1+\exp\left(\Theta_{ij}^{(t)}\right)\right\}\right].
\eeq
Our goal is to find the estimators $\wh{\mT}=(\wh{Z}$; $\{\wh{U}^{(k)}\}_{k=1}^{K}$; $\{\wh{V}^{(t)}\}_{t=1}^T$; $\wh{\mathcal{G}})$ by solving the following optimization problem:
\beq
   \min_{Z,  \{U^{(k)}\}_{k=1}^K, \{V^{(t)}\}_{t=1}^T,\mathcal{G}} \ell_1(\mT).
   \label{obj}
\eeq
To optimize \eqref{obj}, we use the projected gradient descent algorithm \citep{ma2020universal} to estimate the parameters. Specifically, we first compute the first-order derivatives of the log-likelihood function with respect to each parameter, and then iteratively update the parameter estimates. In each iteration, the algorithm moves along the gradient direction with a pre-specified step size, followed by a projection step to rotate the estimates of $U^{(k)}$ as needed.  We define $\sigma(x)=1/(1+e^{-x})$ and summarize the optimization procedure in Algorithm \ref{algorithm 1}.

\begin{algorithm}
\caption{Projected gradient descent method}
\label{algorithm 1}
\begin{algorithmic}[1]
\REQUIRE network adjacency matrices $A^{(1)},...,A^{(T)}$; initial values $Z_{\text{ini}}$, $\{U_{\text{ini}}^{(k)}\}_{k=1}^K$, $\{V_{\text{ini}}^{(t)}\}_{t=1}^T$, initial group label $\mathcal{G}_{\text{ini}}=(g_{1,\text{ini}}, g_{2,\text{ini}},..., g_{T,\text{ini}})$; latent dimension $R_S$, $R_D$; number of groups $K$; step sizes: $\eta_Z$, $\eta_U$, $\eta_V$; number of iterations $M$
\ENSURE  $\wh{Z}=Z_M$; $\{\wh{U}^{(k)}\}_{k=1}^K=\{U^{(k)}_M\}_{k=1}^K$; $\{\wh{V}^{(t)}\}_{t=1}^T=\{V^{(t)}_M\}_{t=1}^T$; group label $\wh{\mathcal{G}}=\{g_{1,M},...,g_{T,M}\}$

\STATE Let $Z_0=Z_{\text{ini}}$;  $\{U^{(k)}_0\}_{k=1}^K=\{U^{(k)}_{\text{ini}}\}_{k=1}^K$; $\{V^{(t)}_0\}_{t=1}^T=\{V^{(t)}_{\text{ini}}\}_{t=1}^T$; $\mathcal{G}_{0}=\mathcal{G}_{\text{ini}}$

\FOR{ $m=0,...,M-1$}

   \STATE $Z_{m+1}=Z_{m}-\eta_Z\nabla\ell_1=Z_{m}+2\eta_Z\left[\sum_{t=1}^{T}\{A^{(t)}-\sigma(\Theta^{(t)})\}\right]Z_m$ 
   \FOR{ $k=1,...,K$ }
   \STATE $U_{m+1}^{(k)}=U_{m}^{(k)}-\eta_U\nabla\ell_1=U_m^{(k)}+2\eta_{U}\sum_{\{t:g_{t,m}=k\}}\{A^{(t)}-\sigma(\Theta^{(t)}_m)\}U^{(k)}_mV^{(t)}_m$
   \STATE  $U^{(k)}_{m+1}=U_{m+1}^{(k)}O^{(k)}_{m+1}$ for an orthogonal matrix $O^{(k)}_{m+1}\in\mR^{R_D\times R_D}$ s.t.   $U^{(k)\top}_{m+1}U^{(k)}_{m+1}= NI_{R_D}$
   \ENDFOR
   
   \FOR{ $t=1,...,T$ }
   \STATE $V_{m+1}^{(t)}=V_{m}^{(t)}-\eta_V\nabla\ell_1=V_m^{(t)}+\eta_{V}\operatorname{diag}\left[U^{(g_{t,m})\top}_m\{A^{(t)}-\sigma(\Theta^{(t)}_m)\}U^{(g_{t,m})}_m\right]$
   \STATE $V_{m+1}^{(t)}=O_{m+1}^{(g_{t,m})\top}V_{(m+1)}^{(t)}O_{m+1}^{(g_{t,m})}$
   \STATE $g_{t,m+1}=\operatorname*{arg\,min}\limits_{g_t\in\{1,...,K\}} \sum_{i,j=1}^{N}\left[A_{ij,m+1}^{(t)}\Theta_{ij,m+1}^{(t)}-\log\{1+\exp(\Theta_{ij,m+1}^{(t)})\}\right]$
  \ENDFOR

\ENDFOR

\end{algorithmic}
\end{algorithm}

Note that Algorithm 1 requires initial estimates $Z_{\text{ini}}$, $\{U_{\text{ini}}^{(k)}\}_{k=1}^K$, $\{V_{\text{ini}}^{(t)}\}_{t=1}^T$, and  $\mathcal{G}_{\text{ini}}$. Given the observed time-varying networks $\{A^{(t)}\}_{t=1}^T$, we first vectorize the upper triangular elements of the $T$ adjacency matrices and then apply $K$-means clustering to obtain the initial group labels $\mathcal{G}_{\text{ini}}$. For the initialization of $Z_{\text{ini}}$, $\{U_{\text{ini}}^{(k)}\}_{k=1}^{K}$, $\{V_{\text{ini}}^{(t)}\}_{t=1}^T$, we consider the simplified version of STANE, which ignores the group structure in the time-varying component. We refer to this model as Simplified STANE, which is specified as   
\begin{equation}
\label{model no group}
\operatorname{logit}(P_{ij}^{(t)})=\Theta^{(t)}=ZZ^\top+U^{(t)}V^{(t)}U^{(t)\top},
\end{equation}
and the negative $\log$-likelihood function is 
\[
\ell_2\left(Z; \{U^{(t)}\}_{t=1}^{T}; \{V^{(t)}\}_{t=1}^T;\mathcal{G}\right)=-\sum_{t=1}^{T}\sum_{i=1}^{N}\sum_{j=1}^{N} \left[A_{ij}^{(t)}\Theta_{ij}^{(t)}-\log\left\{1+\exp\left(\Theta_{ij}^{(t)}\right)\right\}\right].
\]
The Simplified STANE model is also optimized using a projected gradient descent method \citep{ma2020universal}. Further details on the initialization and the projected gradient algorithm are provided in Appendix B.

\subsubsection{Estimation of Sparse STANE}
\label{variable selection}
To handle the presence of all-zero rows in $U^{(k)}$ and to effectively select the nonzero dynamic components, we introduce a regularization term into the objective function. Specifically, for the Sparse STANE model, we propose the following regularized optimization problem:
\begin{gather}
    \min_{Z,  \{U^{(k)}\}_{k=1}^K, \{V^{(t)}\}_{t=1}^T,\mathcal{G}} Q\left(\mT\right) :=
    \frac{1}{N^2}\ell_1\left( \mT \right)  +\sum_{k=1}^{K}\sum_{i=1}^{N}\rho(\Vert U^{(k)}_{i\cdot}\Vert_2; \mu, \gamma),
    \label{model3}
\end{gather}
where $\rho(\cdot)$ is a minimax concave penalty (MCP), $\mu$ and $\gamma$ are two tuning parameters. For any $t\geq 0$, $\rho(t) =\mu\int_{0}^{t}(1-x/(\mu\gamma))_{+}dx$. The tuning parameter $\mu$ is selected by $\BIC$ and $\gamma $ is set to 3 by default.  Additionally, the latent dimensions $R_S$ and $R_D$, as well as the number of groups $K$, are also chosen using the BIC criterion. To determine the optimal values of $R_S$ and $R_D$, we initially set $K = T$, and define the $\BIC_1$ criterion as follows
\beq
  \BIC_1(R_S, R_D)= \ell_1(\mT)+\log(N^2T)\left\{(R_S+TR_D)N+TR_D\right\}.
\eeq
Given $R_S$, $R_D$, we select $K$ and $\mu$ by
\beq
  \BIC_2(K, \mu)= \ell_1(\mT)+C_0\log(N^2T)\sum_{k=1}^K \mathcal{H}_k,
\eeq
where $\mathcal{H}_k$ is the number of non-zero rows of $U^{(k)}$ for $k=1,...,K$ and $C_0$ is a constant. We set $C_0=0.1$ in our numerical experiments.
The optimization procedure for problem \eqref{model3} is similar to that of Algorithm \ref{algorithm 1},  except that we update $U^{(k)}$ using the penalized version to account for the regularization. Specifically, during each iteration, the soft-thresholding operator is applied to each row of $U^{(k)}$ for $k=1,...,K$, which can be expressed as, 
\beq
  \mS_{\mu, \gamma}(U_{i\cdot}^{(k)})=
  \begin{cases}
  \dfrac{(1-\mu/\Vert U_{i\cdot}^{(k)}\Vert_2)_+U_{i\cdot}^{(k)}}{1-1/\gamma},~~~\Vert U_{i\cdot}^{(k)}\Vert_2<\mu\gamma \\
U_{i\cdot}^{(k)},~~~~~~~~~~~~~~~~~~~~~~~~~~\Vert U_{i\cdot}^{(k)}\Vert_2\geq\mu\gamma  .
  \end{cases}
\eeq

\section{Theoretical guarantees}
\label{Theoretical guarantees}

In this section, we establish the theoretical properties of STANE and Sparse STANE. We begin by deriving an upper bound on the estimation error of $\bm{\Theta}$, and then proceed to provide upper bounds for the estimation errors of $Z$ and $\{U^{(k)}\}_{k=1}^K$.
Finally, we prove the variable selection consistency of $\{U^{(k)}\}_{k=1}^K$. To facilitate our theoretical analysis, we first introduce some notation. 
Let the true parameters be denoted by $\mT^*=(Z^*; \{U^{(k)*}\}_{k=1}^{K}; \{V^{(t)*}\}_{t=1}^T;\mathcal{G}^*)$, where $\mathcal{G}^*=\{g_1^*,...,g_T^*\}$. Define $S^*=Z^*Z^{*\top}$, $D^{(t)*}=U^{(g_t^*)*}V^{(t)*}U^{(g_t^*)*\top}$, and further let $\bm{S}^*=[S^*;...;S^*]$, $\bm{D}^*=[D^{(1)*};...;D^{(T)*}]$, and $\bm{\Theta}^*=\bm{S}^*+\bm{D}^*$. For each $k=1,...,K$, define the support of $U^{(k)*}$ as $H^{(k)}=\{i: U_{i\cdot}^{(k)*\top}\neq \bm0_{R_D}\}$. We denote the oracle estimators by $\wh{\mT}^{or}=(\wh{Z}^{or}; \{\wh{U}^{(k)or}\}_{k=1}^K; \{\wh{V}^{(t)or}\}_{t=1}^T; \wh{\mathcal{G}}^{or})$, where $\wh{\mathcal{G}}^{or}=\{\wh{g}_{1}^{or},...,\wh{g}_{T}^{or}\}$. The oracle estimators correspond to the estimators obtained when the true underlying sparse structure is known. Specifically, they are the solutions to the following optimization problem:
\[
\underset{Z, \{U^{(k)}\}_{k=1}^K,\{V^{(t)}\}_{t=1}^T,\mathcal{G}}{\operatorname{arg~min}}~
\ell_1(\mT), ~~~~~\text{s.t.}~U^{(k)\top}_{i\cdot}=\bm{0}_{R_D} ~\text{for}~ i\notin H^{(k)} ~\text{and}~ k=1,...,K.
\]


The corresponding oracle estimators are given by $\wh{\Theta}^{(t)or}=\wh{Z}^{or}\wh{Z}^{or\top}+\wh{U}^{(\wh{g}_t^{or})or}\wh{V}^{(t)or}\wh{U}^{(\wh{g}_t^{or})or\top}=\wh{S}^{or}+\wh{D}^{(t)or}$, where $\wh{S}^{or} = \wh{Z}^{or}\wh{Z}^{or\top}$, and $\wh{D}^{(t)or} = \wh{U}^{(\wh{g}_t^{or})or}\wh{V}^{(t)or}\wh{U}^{(\wh{g}_t^{or})or\top}$. We further denote $\wh{\bm{\Theta}}^{or}=[\wh{\Theta}^{(1)or};...;\wh{\Theta}^{(T)or}]$, $\wh{\bm{S}}^{or}=[\wh{S}^{or};...;\wh{S}^{or}]$ and $\wh{\bm{D}}^{or}=[\wh{D}^{(1)or};...;\wh{D}^{(T)or}]$ as the oracle estimators of $\bm{\Theta}$, $\bm{S}$ and $\bm{D}$.  Before presenting our theoretical results, we need to establish
some conditions.

\begin{enumerate}
\item[(C1)] \label{con:C1} There exist positive constants $\tau_1$ and $\tau_2$ such that $\Vert \Theta^{(t)}\Vert_{\operatorname{max}}\leq \tau_1$, $\Vert V^{(t)}\Vert_{\max}\leq \tau_2$ for $t=1,...,T$.
\item [(C2)]  \label{con:C2} The matrices $\{D^{(t)}\}_{t=1}^{T}$  satisfy the centering condition  $\sum_{t=1}^T D^{(t)}=\bm{0}_{N\times N}$.
\end{enumerate}

Note that the boundedness conditions on $\{\Theta^{(t)}\}_{t=1}^{T}$ and $\{V^{(t)}\}_{t=1}^T$ are imposed purely for theoretical analysis, which ensures latent positions are not excessively large, preventing distortion in the modeling process. Similar conditions have been adopted in \cite{ma2020universal} and \cite{lyu2023latent}. The constraint $\sum_{t=1}^T D^{(t)} = \bm{0}_{N\times N}$ serves as a technical condition to guarantee that $\|\wh{\bm{D}}^{or} - \bm{D}^*\|_F^2 \leq \|\wh{\bm{\Theta}}^{or} - \bm{\Theta}^*\|_F^2$, making temporal changes relative and promoting consistency and interpretability across time periods. Under conditions (C1) and (C2), we establish an upper bound on the estimation error of $\wh{\bm{\Theta}}^{or}$ in the following Theorem \ref{thm1}.

\bet
\label{thm1}
Under conditions (C1)-(C2), there exist constants $c_1$ and $c_2$ such that, with probability at least $1-T \exp(-c_1N)-\exp\{-c_2(2N+T)\}$, we have
\beq
\label{eq in thm1}
\Vert\wh{\bm{\Theta}}^{or}-\bm{\Theta}^*\Vert_F^2 \leq C_1NR_S+C_2K^2R_D^2(2N+T),
\eeq
where $C_1$ and $C_2$ are constants depending on $\tau_1$.
\eet

As $ N \to \infty $, the probability in Theorem \ref{thm1} converges to $1$. Besides, when $T$, $R_S$, $R_D$, and $K$ are fixed, Theorem \ref{thm1} indicates that $\Vert\wh{\bm{\Theta}}^{or}-\bm{\Theta}^*\Vert_F^2/(N^2T)=O_p(1/N)$, with probability at least $1-T \exp(-c_1N)-\exp\{-c_2(2N+T)\}$. The upper bound is consistent with the result in \citep{zhang2022joint} and the proof is given in the Appendix A.2. Next, we establish the upper 
bounds on the estimation errors of $\wh{Z}^{or}$ and $\{\wh{U}^{(k)or}\}_{k=1}^K$. To proceed, we introduce the following condition (C3).

\begin{enumerate}
    \item[(C3)]\label{con:C3} 
    For the true parameters $\{V^{(t)*}\}_{t=1}^T$, there exists a constant $\kappa >0$ such that $\Vert V^{(t)*}\Vert_{\min}\geq\kappa$ for $t=1,...,T$, where $\kappa\leq \tau_2$, and $\tau_2$ is defined in condition (C1).
\end{enumerate}
Condition (C3) further limits the minimum absolute value of $\{V^{(t)*}\}_{t=1}^T$ to be greater than a constant $\kappa$ to ensure time-varying components in the latent space are strong enough to stand out from noise. Similar constraint can be found in \citep{pmlr-v119-zhang20aa}. Under this condition, we next present the error bounds for $\wh{Z}^{or}$ and $\{\wh{U}^{(k)or}\}_{k=1}^K$.

\begin{corollary}
\label{coro 1}
We assume that there exists a constant $\delta>0$ such that $T\leq \delta N$. Then, under conditions (C1)-(C3), we have 
\beq
\Vert\wh{Z}^{or}\wh{Z}^{or\top}-Z^*Z^{*\top} \Vert^2_F \leq C_1NR_S/T+C_2K^2R_D^2(2N+T)/T,
\eeq
and for $k=1,...,K$,
\beq
\min_{O:OO^\top=O^\top O=I_{R_D}}\left\{ \Vert\wh{U}^{(k)or}-U^{(k)*}O\Vert_F^2\right\}\leq 8(n_k\kappa^2)^{-1}(C_1R_S+\wt{C}_2K^2R_D^2)
,
\eeq
with probability at least $1-T \exp(-c_1N)-\exp\{-c_2(2N+T)\}$, where $n_k$ is the number of time points in group $k$,   $\wt{C}_2=C_2(2+\delta)$ , $C_1$, $C_2$, $c_1$, $c_2$ are defined in Theorem \ref{thm1}, and $\kappa$ is defined in condition (C3).

\end{corollary}

Corollary \ref{coro 1} shows that the upper bounds on the estimation errors of $Z$ and $\{U^{(k)}\}_{k=1}^{K}$ decrease at rates inversely proportional to $T$ and $n_k$,  respectively, with probability approaching 1 as the number of nodes tends to infinity. The proof of Corollary \ref{coro 1} is presented in the Appendix A.3. Lastly, we present the theoretical result on the accuracy of variable selection, which requires the following additional conditions. 

\begin{enumerate}
 \item [(C4)]\label{con:C4}
    For $k=1,...,K$,  assume that $\min_{i\in H^{(k)}}\Vert U^{(k)*}_{i\cdot}\Vert_2>\mu\gamma+\{8(n_k\kappa^2)^{-1}(C_1R_S+\wt{C}_2K^2R_D^2)\}^{\frac{1}{2}}$, where $C_1$, $\wt{C}_2$ and $\kappa$ are  defined in Theorem \ref{thm1}, Corollary \ref{coro 1} and condition (C3) respectively. Furthermore, there exist a constant $C_\mu<1$ such that $\{8(n_k\kappa^2)^{-1}(C_1R_S+\wt{C}_2K^2R_D^2)\}^{\frac{1}{2}}<C_\mu\mu\gamma$ for $k=1,...,K$.

\item [(C5)]\label{con:C6} 
    Define $d>0$, such that $N\max_t\Vert P^{(t)}\Vert_{\max}\leq d$. Then there exists a constant $c_3\in(0,1)$ such that $c_3\mu N^2T^{-1}(1-C_\mu)\{2\tau_2(\sqrt{N}+C_\mu\mu\gamma)\}^{-1}>\sqrt{d}$, and 
    $\mu N^2(1-c_3)(1-C_\mu)\{\tau_2(\sqrt{N}+C_\mu\mu\gamma)\}^{-1} \geq  \sqrt{T}\{C_1NR_S+C_2K^2R_D^2(2N+T)\}^\frac{1}{2}$, where $C_1$, $C_2$ are defined in Theorem \ref{thm1}, $\tau_2$ is defined in condition (C1) and $C_\mu$ is defined in condition (C4).
\end{enumerate}

Condition (C4) imposes a minimal signal strength requirement on the nonzero rows $\{U^{(k)*}_{i\cdot}\}_{k=1}^K$ for $i\in H^{(k)}$, which is crucial for ensuring that these non-zero rows can be detected above the noise level.  Condition (C5) provides a lower bound on $\mu$, ensuring that $\mu$ is large enough to suppress false positives while retaining true signals. Then we give the consistency of selecting non zero time-varying latent positions.


\bet
\label{thm2}
Under conditions (C1)-(C5), there exist a local minimizer of \eqref{model3} $\wh{\mT}^p=(\wh{Z}^p;\{\wh{U}^{(k)p}\}_{k=1}^K;\{\wh{V}^{(t)p}\}_{t=1}^T;\wh{\mathcal{G}}^p)$, such that  $\wh{\mT}^p=\wh{\mT}^{or}$ with probability tending to 1 when N goes to infinity. 
Moreover, there exist constants $r_{t},~t=1,...,T$ such that $\wh{U}^{(k)p\top}_{i\cdot}=\bm{0}_{R_D}$ for $i\notin H^{(k)}$ and $k=1,...,K$ with probability at least
\begin{equation}
1-\sum_{t=1}^{T}N^{-r_{t}}-2T\exp(-c_1N)-2\exp\{-c_2(2N+T)\}.
\label{ex14}
\end{equation}

\eet

Theorem \ref{thm2} indicates that non-zero elements of $\{U^{(k)*}\}_{k=1}^K$ can be identified with a probability tending to 1 when $N$ goes to infinity. The proof of Theorem \ref{thm2} is provided in the Appendix A.4.

\section{Simulation studies}
\label{sim}

In this section, we demonstrate the efficiency and effectiveness of the proposed model through various simulation studies. Specifically, we assess the estimation accuracy of the parameters $Z$, $\{U^{(k)}
\}_{k=1}^K$, $\{V^{(t)}\}_{t=1}^T$, $\{P^{(t)}\}_{t=1}^T$, as well as the model's ability to correctly classify group memberships. Estimation errors are evaluated under different settings and compared across several methods.

Let $\wh{\mT}=(\wh{Z};\{\wh{U}^{(k)}\}_{k=1}^K;\{\wh{V}^{(t)}\}_{t=1}^T;\wh{\mathcal{G}})$ denote the estimators obtained from optimization problem \eqref{obj} and \eqref{model3} with slight abuse of notation.
We evaluate the estimation errors using the following relative error metrics:
$\text{Z.error}=\Vert Z^*Z^{*\top} - \wh{Z}\wh{Z}^\top\Vert_F^2/\Vert Z^*Z^{*\top}\Vert_F^2$,  $\text{U.error}=\sum_{k=1}^K\Vert U^{(k)*}U^{(k)*\top} - \wh{U}^{(k)}\wh{U}^{(k)\top}\Vert_F^2/\sum_{k=1}^K\Vert U^{(k)*}U^{(k)*\top}\Vert_F^2$ and $\text{V.error}=\sum_{t=1}^T\Vert \wh{V}^{(t)}-V^{(t)*}O^{(t)}\Vert_F^2/\sum_{t=1}^{T}\Vert V^{(t)*}\Vert_F^2$, 
where $O^{(t)}=\operatorname{arg min}_{O^{(t)}O^{(t)\top}=I_{R_D}}
\Vert \wh{V}^{(t)}-V^{(t)*}O^{(t)}\Vert_F^2$.
The optimal $O^{(t)}$ can be obtained by singular value decomposition (SVD).  Additionally, the estimation error for $\{P^{(t)}\}_{t=1}^T$ is evaluated by $\text{P.error}=\sum_{t=1}^{T}\Vert P^{(t)*} - \wh{P}^{(t)}\Vert_F^2/\sum_{t=1}^{T}\Vert P^{(t)*}\Vert_F^2$. We also evaluate the quality of group classification using the normalized mutual information (NMI) index, a widely used metric for clustering performance.

We investigate two distinct scenarios: (1) $\{U^{(k)}\}_{k=1}^K$ are dense; (2) $\{U^{(k)}\}_{k=1}^K$ are sparse. For each scenario, we explore four specific cases to evaluate the impact of key factors on estimation performance.
These factors include the network size $N$, the number of time points $T$, the number of groups $K$  and the dimension of the latent positions $Z$ and $\{U^{(k)}\}_{k=1}^K$. By varying one aspect while keeping the other three fixed, we observe how the estimation error changes and compare the results with different methods. For each parameter setting, we conduct 100 independent replications to ensure statistical reliability.

\subsection{$\{U^{(k)}\}_{k=1}^K$ are dense}
When $\{U^{(k)}\}_{k=1}^K$ are dense, the true model parameters are generated as follows. First, generate $Z_{ij}^*$ from standard normal distribution $\mN(0,1)$ independently for $i=1,...,N,~j=1,...,R_S$. Second, for $k=1,...,K$, generate $U^{(k)*}_{ij}$ from Uniform $(0.3,0.8)$ independently for $i=1,...,N$, $j=1,...,R_D$, and rotate $U^{(k)*}$ s.t. $U^{(k)*\top}U^{(k)*}=N I_{R_D}$. Third, for the diagonal matrices $\{ V^{(t)*}\}_{t=1}^T$,  generate $V^{(t)*}_{ii}$ from Uniform$(2.5,3)$ independently for $t=1,...,T$, $i=1,...,R_D$.
After setting the true parameters, we generate $\{A^{(t)}\}_{t=1}^T$ based on model \eqref{model1}.

For initial value of Algorithm 2 presented in Appendix A2, we generate each entry of $Z_{\text{ini},0}$ independently from $\mN(0,1)$. For $U_{\text{ini},0}^{(t)}$, each entry is generated independently from Uniform$(0.3,0.8)$, and then transform $U_{\text{ini},0}^{(t)}$ such that $U_{\text{ini},0}^{(t)\top}U_{\text{ini},0}^{(t)}=NI_{R_D}$ for $t=1,...,T$. Additionally, we generate diagonal matrix $V^{(t)}_{\text{ini},0}$ by letting each diagonal entry follow Uniform$(2.5,3)$ independently for $t=1,...,T$. The step sizes are set as $\eta_Z=\eta/\Vert Z\Vert_2^2$, $\eta_U=\eta/\Vert U^{(1)}\Vert_2^2$ and  $\eta_V=\eta/(NT\Vert U^{(1)}\Vert_2^2)$ where  $\eta$ is a small positive constant.

We then examine the four cases described above. For each case, we compare the estimation results with several widely-used methods designed to analyze the shared and specific structures of multilayer networks, recognizing that dynamic networks can be viewed as a special case of multilayer networks. Specifically, we consider the MultiNeSS model proposed by \cite{macdonald2022latent} and the M-GRAF algorithm introduced by \cite{wang2019common}. In addition, we include the Simplified STANE model, as introduced in Section~\ref{parameter estimation}, and refer to this method as Sim-STANE for notational simplicity.

\textsc{Case 1} (\textsc{Network Size}).
To demonstrate the effect of the network size on estimation performance, we fix $T=20$, $K=3$, $R_S=2$, $R_D=3$, and vary network size by $N=\{200,400,800\}$. For each group, we ensure it has at least  5 time points, and randomly assign group labels to the remaining time points.
The estimation results are presented in Table \ref{table1}. 
We can see that, the estimation error of STANE decreases as the network size increases . Additionally, the NMI index remain unchanged at 1, indicating perfect accuracy in group label estimation. Moreover, the relative error of all parameters are lower than Sim-STANE indicating the necessity of considering the group structure of dynamic parts.  In terms of probability matrix estimation, the estimation error of all methods is decreasing as $N$ increases.
Additionally, STANE outperforms all other methods in probability matrix estimation.

\begin{table}[t]
\centering
\setlength{\tabcolsep}{2pt}
\captionsetup{skip=8pt}

    \caption{The mean and standard deviation (standard deviations are in brackets) of estimation error of $Z$, $\{U^{(k)}\}_{k=1}^K$, $\{V^{(t)}\}_{t=1}^T$, $\{P^{(t)}\}_{t=1}^T$ and NMI index, when $\{U^{(k)}\}_{k=1}^K$ are dense and $T=20$, $K=3$, $R_S=2$, $R_D=3$.}
\resizebox{\textwidth}{!}{
\begin{tabular}{cccccccccccc}
\toprule
& \multicolumn{2}{c}{Z.error} 
& \multicolumn{2}{c}{U.error} 
& \multicolumn{2}{c}{V.error} 
& \multicolumn{4}{c}{P.error}
& \multicolumn{1}{c}{NMI} \\
\cmidrule(r){2-3} \cmidrule(r){4-5} \cmidrule(r){6-7} \cmidrule(r){8-11} \cmidrule(r){12-12} 

\shortstack{$N$\\~} &  \shortstack{Sim-\\STANE} & \shortstack{STANE\\~} & \shortstack{Sim-\\STANE} & \shortstack{STANE\\~} & \shortstack{Sim-\\STANE} & \shortstack{STANE\\~} & \shortstack{MultiNeSS\\~} & \shortstack{M-GRAF\\~} & \shortstack{Sim-\\STANE} & \shortstack{STANE\\~} & \shortstack{STANE\\~} \\
\midrule
\multirow{2}{*}{200} 
& 0.0404 & 0.0276 & 0.0368 & 0.0128 & 0.0140 & 0.0119 & 0.0056 & 0.0250 & 0.0030 & 0.0010 & 1.0000 \\
& (0.0036) & (0.0034) & (0.0022) & (0.0023) & (0.0011) & (0.0008) & (0.0029) & (0.0004) & (0.0001) & (0.0000) & (0.0000) \\ [3pt]

\multirow{2}{*}{400} 
& 0.0267 & 0.0147 & 0.0295 & 0.0070 & 0.0133 & 0.0092 & 0.0038 & 0.0180 & 0.0024 & 0.0006 & 1.0000 \\
& (0.0015) & (0.0011) & (0.0014) & (0.0006) & (0.0009) & (0.0008) & (0.0023) & (0.0001) & (0.0002) & (0.0000) & (0.0000) \\ [3pt]

\multirow{2}{*}{800} 
& 0.0209 & 0.0101 & 0.0250 & 0.0062 & 0.0114 & 0.0073 & 0.0014 & 0.0077 & 0.0020 & 0.0004 & 1.0000 \\
& (0.0011) & (0.0005) & (0.0012) & (0.0002) & (0.0007) & (0.0006) & (0.0004) & (0.0000) & (0.0001) & (0.0000) & (0.0000) \\
\bottomrule
\end{tabular}
}
\label{table1}
\end{table}

\begin{table}[t]
\centering
\setlength{\tabcolsep}{2pt}

\captionsetup{skip=8pt}
    \caption{The mean and standard deviation (standard deviations are in brackets) of estimation error of $Z$, $\{U^{(k)}\}_{k=1}^K$, $\{V^{(t)}\}_{t=1}^T$, $\{P^{(t)}\}_{t=1}^T$ and NMI index, when $\{U^{(k)}\}_{k=1}^K$ are dense and $N=200$, $K=3$, $R_S=2$, $R_D=3$.}
\resizebox{\textwidth}{!}{
\begin{tabular}{cccccccccccc}
\toprule

& \multicolumn{2}{c}{Z.error} 
& \multicolumn{2}{c}{U.error} 
& \multicolumn{2}{c}{V.error} 
& \multicolumn{4}{c}{P.error}
& \multicolumn{1}{c}{NMI} \\
\cmidrule(r){2-3} \cmidrule(r){4-5} \cmidrule(r){6-7}  \cmidrule(r){8-11} \cmidrule(r){12-12}
\shortstack{$T$\\~} & \shortstack{Sim-\\STANE} & \shortstack{STANE\\~} & \shortstack{Sim-\\STANE} & \shortstack{STANE\\~} & \shortstack{Sim-\\STANE} & \shortstack{STANE\\~} & \shortstack{MultiNeSS\\~} & \shortstack{M-GRAF\\~} & \shortstack{Sim-\\STANE} & \shortstack{STANE\\~} & \shortstack{STANE\\~} \\
\midrule
\multirow{2}{*}{20}  
& 0.0404 & 0.0276 & 0.0368 & 0.0128 & 0.0140 & 0.0119 & 0.0056 & 0.0250 & 0.0030 & 0.0010 & 1.0000 \\

& (0.0036) & (0.0034) & (0.0022) & (0.0023) & (0.0011) & (0.0008) & (0.0029) & (0.0004) & (0.0001) & (0.0000) & (0.0000) \\ [3pt]

\multirow{2}{*}{25} 
& 0.0345 & 0.0232 & 0.0357 & 0.0105 & 0.0120 & 0.0119 & 0.0056 & 0.0231 & 0.0030 & 0.0008 & 1.0000 \\
& (0.0027) & (0.0020) & (0.0013) & (0.0012) & (0.0012) & (0.0008) & (0.0029) & (0.0003) & (0.0001) & (0.0000) & (0.0000) \\ [3pt]

\multirow{2}{*}{30} 
& 0.0319 & 0.0201 & 0.0355 & 0.0092 & 0.0119 & 0.0114 & 0.0056 & 0.0218 & 0.0030 & 0.0007 & 1.0000 \\
& (0.0027) & (0.0015) & (0.0010) & (0.0010) & (0.0010) & (0.0007) & (0.0029) & (0.0003) & (0.0001) & (0.0000) & (0.0000) \\
\bottomrule
\end{tabular}
}
\label{table2}
\end{table}

\textsc{Case 2} (\textsc{Number of Time Points}).
To examine how the estimation error changes with an increasing number of time points $T$ , we fix $N=200$, $K=3$, $R_S=2$, $R_D=3$ and vary $T=\{20,25,30\}$. For each $T$, we ensure it has at least 5, 6, 7 time points in each group respectively, and randomly assign group labels to the remaining time points. Table \ref{table2} shows the estimation results. It can be seen that, as the number of time points increases, the relative error of all parameters based on STANE decreases. NMI index still equal to 1 indicating the stability in the estimation of group label. As $T$ increases, the estimation error of the probability matrices $\{P^{(t)}\}_{t=1}^T$ for MultiNeSS, M-GRAF and Sim-STANE decreases and STANE outperforms all methods in parameter estimation and probability matrix estimation.

\textsc{Case 3} (\textsc{Number of Groups}).
To investigate the effect of $K$, we vary $K=\{3,4,5\}$ while fixing the other parameters as $N=400$, $T=25$, $R_S=2$, $R_D=3$. For each $K$, we ensure that each group contains at least 6, 5, 4 time points respectively and randomly assign group labels to the remaining time points. The simulation results are demonstrated in Table \ref{table3}. We observe that as $K$ increases, the number of parameters that need to be estimated for $\{U^{(k)}\}_{k=1}^K$ grows, 
leading to a rise in estimation error of the proposed model. However, the value of $K$ does not affect the other methods, since these methods do not model the group structure. The estimation errors of the other three methods remain relatively constant across different value of $K$. Nevertheless, STANE still has the best performance.

\textsc{Case 4} (\textsc{Dimension of Latent Positions}).
Lastly, we consider how estimation error change with respect to dimension of latent positions. We vary $R_S=\{2,4,6\}$, $R_D=\{3,5,7\}$, and fix the other parameters as $N=200$, $T=20$, $K=3$. Similar to the previous case, we also ensure that each group contains at least 5 time points and randomly assign group label to the remaining time points. The estimation results are shown in Table \ref{table4}. It can be seen that estimation results of all parameters become worse as the dimensions of latent positions get larger, due to the larger number of parameters that need to be estimated. Under this case, STANE still has the best estimating performance.

\begin{table}[t]
\centering
\setlength{\tabcolsep}{2pt}
\captionsetup{skip=8pt}
    \caption{The mean and standard deviation (standard deviations are in brackets) of estimation error of $Z$, $\{U^{(k)}\}_{k=1}^K$, $\{V^{(t)}\}_{t=1}^T$, $\{P^{(t)}\}_{t=1}^T$ and NMI index, when $\{U^{(k)}\}_{k=1}^K$ are dense and $N=400$, $T=25$, $R_S=2$, $R_D=3$.}
\resizebox{\textwidth}{!}{
\begin{tabular}{cccccccccccc}
\toprule

& \multicolumn{2}{c}{Z.error} 
& \multicolumn{2}{c}{U.error} 
& \multicolumn{2}{c}{V.error} 
& \multicolumn{4}{c}{P.error}
& \multicolumn{1}{c}{NMI} \\
\cmidrule(r){2-3} \cmidrule(r){4-5} \cmidrule(r){6-7}  \cmidrule(r){8-11} \cmidrule(r){12-12}
\shortstack{$K$\\~} &  \shortstack{Sim-\\STANE} & \shortstack{STANE\\~} & \shortstack{Sim-\\STANE} & \shortstack{STANE\\~} & \shortstack{Sim-\\STANE} & \shortstack{STANE\\~} & \shortstack{MultiNeSS\\~} & \shortstack{M-GRAF\\~} & \shortstack{Sim-\\STANE} & \shortstack{STANE\\~} & \shortstack{STANE\\~} \\
\midrule
\multirow{2}{*}{3} 
& 0.0285 & 0.0112 & 0.0281 & 0.0066 & 0.0124 & 0.0085 & 0.0038 & 0.0160 & 0.0023 & 0.0005 & 1.0000 \\
& (0.0016) & (0.0009) & (0.0010) & (0.0004) & (0.0009) & (0.0006) & (0.0023) & (0.0001) & (0.0001) & (0.0000) & (0.0000) \\ [3pt]

\multirow{2}{*}{4} 
& 0.0283 & 0.0124 & 0.0285 & 0.0079 & 0.0122 & 0.0090 & 0.0023 & 0.0161 & 0.0023 & 0.0005 & 1.0000 \\
& (0.0017) & (0.0011) & (0.0010) & (0.0010) & (0.0009) & (0.0007) & (0.0013) & (0.0001) & (0.0001) & (0.0000) & (0.0000) \\ [3pt]

\multirow{2}{*}{5} 
& 0.0281 & 0.0133 & 0.0285 & 0.0085 & 0.0120 & 0.0090 & 0.0020 & 0.0156 & 0.0023 & 0.0006 & 1.0000 \\
& (0.0017) & (0.0014) & (0.0010) & (0.0007) & (0.0010) & (0.0007) & (0.0015) & (0.0001) & (0.0001) & (0.0000) & (0.0000) \\
\bottomrule
\end{tabular}
}
\label{table3}
\end{table}

\begin{table}[t]
\centering
\setlength{\tabcolsep}{1.5pt}
\captionsetup{skip=8pt}
    \caption{The mean and standard deviation (standard deviations are in brackets) of estimation error of $Z$, $\{U^{(k)}\}_{k=1}^K$, $\{V^{(t)}\}_{t=1}^T$, $\{P^{(t)}\}_{t=1}^T$ and NMI index, when $\{U^{(k)}\}_{k=1}^K$ are dense and $N=200$, $T=20$, $K=3$.}
\resizebox{\textwidth}{!}{
\begin{tabular}{cccccccccccccc}
\toprule

& \multicolumn{2}{c}{Z.error} 
& \multicolumn{2}{c}{U.error} 
& \multicolumn{2}{c}{V.error} 
& \multicolumn{4}{c}{P.error}
& \multicolumn{1}{c}{NMI} \\
\cmidrule(r){2-3} \cmidrule(r){4-5} \cmidrule(r){6-7} \cmidrule(r){8-11} \cmidrule(r){12-12}
\shortstack{$R_S,R_D$\\~} &  \shortstack{Sim-\\STANE} & \shortstack{STANE\\~} & \shortstack{Sim-\\STANE} & \shortstack{STANE\\~} & \shortstack{Sim-\\STANE} & \shortstack{STANE\\~} & \shortstack{MultiNeSS\\~} & \shortstack{M-GRAF\\~} & \shortstack{Sim-\\STANE} & \shortstack{STANE\\~} & \shortstack{STANE\\~} \\
\midrule
\multirow{2}{*}{2, ~3} 
& 0.0404 & 0.0276 & 0.0368 & 0.0128 & 0.0140 & 0.0119 & 0.0056 & 0.0250 & 0.0030 & 0.0010 & 1.0000 \\
& (0.0036) & (0.0034) & (0.0022) & (0.0023) & (0.0011) & (0.0008) & (0.0029) & (0.0004) & (0.0001) & (0.0000) & (0.0000) \\ [3pt]

\multirow{2}{*}{4, ~5} 
& 0.0932 & 0.0603 & 0.0428 & 0.0137 & 0.0318 & 0.0321 & 0.0048 & 0.0279 & 0.0033 & 0.0015 & 1.0000 \\
& (0.0043) & (0.0019) & (0.0011) & (0.0009) & (0.0022) & (0.0014) & (0.0005) & (0.0002) & (0.0001) & (0.0000) & (0.0000) \\ [3pt]

\multirow{2}{*}{6, ~7} 
& 0.1450 & 0.1220 & 0.0508 & 0.0138 & 0.0524 & 0.0693 & 0.0064 & 0.0265 & 0.0038 & 0.0026 & 1.0000 \\
& (0.0286) & (0.0213) & (0.0009) & (0.0007) & (0.0032) & (0.0033) & (0.0002) & (0.0002) & (0.0015) & (0.0013) & (0.0000) \\
\bottomrule
\end{tabular}
}
\label{table4}
\end{table}

\subsection{$\{U^{(k)}\}_{k=1}^K$ are sparse}
We investigate the estimation performance of the Sparse-STANE model when deal with the sparse change of dynamic networks.  
The true parameters are generated similarly to the previous subsection except $\{U^{(k)}\}_{k=1}^K$. Specifically, we randomly select $s_0$ proportion of the rows of $\{U^{(k)}\}_{k=1}^K$ to be zero, and we set $s_0$ to be 0.3. The generation of initial values and the step sizes are the same as in the previous scenario. We also investigate the estimation performance of the Sparse-STANE model when varying the network size $N$, the number of time points $T$, the number of groups $K$, the dimension of latent positions $R_S$, $R_D$, following the same setup as the previous subsection. We denote Sparse-STANE as Spa-STANE for notational simplicity. Additionally, we include one more method for comparison, which is Sparse Simplified STANE and we denote this as Spa-Sim-STANE. Spa-Sim-STANE is the extension of Simplified STANE model to allow for partially time invariant interactions. That is $U_{i\cdot}^{(t)\top}=\bm{0}_{R_D}$ for some $i\in\{1,...,N\}$. We  then optimize the following optimization problem:
\[
\underset{Z,\{U^{(t)}\}_{t=1}^T,\{V^{(t)}\}_{t=1}^T,\mathcal{G}}{\min}
\frac{1}{N^2}\ell_2\left( Z;\{U^{(t)}\}_{t=1}^T;\{V^{(t)}\}_{t=1}^T;\mathcal{G} \right)  +\sum_{t=1}^{T}\sum_{i=1}^{N}\rho(\Vert U^{(t)}_{i\cdot}\Vert_2; \mu, \gamma).
\]
The estimation error of the probability matrices $\{P^{(t)}\}_{t=1}^T$ for the four cases are presented in Table \ref{table5}. We can see that the Spa-STANE has the best performance among all other methods. Specifically, the better performance compared to STANE indicates that parameter estimation can be improved through the penalized approach when $\{U^{(k)}\}_{k=1}^K$ are sparse. The more estimation results of other parameters are presented in the Appendix C.

To assess the variable selection accuracy of Spa-STANE and Spa-Sim-STANE, we report the True Positive Rate (TPR) and False Positive Rate (FPR) under different sparsity levels $s_0$ in Table \ref{table6}. The results indicate that variable selection accuracy improves as $N$ increases. Moreover, Spa-STANE outperforms Spa-Sim-STANE, highlighting the importance of incorporating group structure when modeling sparse changes in dynamic networks.
\begin{table}[htbp]
\centering
\footnotesize  
\setlength{\tabcolsep}{5pt} 

\captionsetup{skip=8pt}

    \caption{The mean and standard deviation (standard deviations are in brackets) of the  estimation error of $\{P^{(t)}\}_{t=1}^T$ and NMI index in different cases, when $\{U^{(k)}\}_{k=1}^K$ are sparse. }
\resizebox{\textwidth}{!}{
\begin{tabular}{ccccccccc}
\toprule
& \multicolumn{6}{c}{P.error} 
& \multicolumn{2}{c}{NMI} \\

\cmidrule(r){2-7}  \cmidrule(r){8-9}
 & \shortstack{MultiNeSS\\~} & \shortstack{M-GRAF\\~} & \shortstack{Sim-\\STANE}  & \shortstack{Spa-Sim-\\STANE}  & \shortstack{STANE\\~} & \shortstack{Spa-\\STANE}  & \shortstack{STANE\\~} & \shortstack{Spa-\\STANE}  \\
\midrule

$N$ & \multicolumn{8}{c}{$T=20$, $K=3$, $R_D=2$, $R_S=3$} \\ [3pt]

\multirow{2}{*}{200} 
& 0.0053 & 0.0298 & 0.0059 & 0.0033 & 0.0013 & 0.0011 & 1.0000 & 1.0000  \\
& (0.0002) & (0.0003) & (0.0002) & (0.0002) & (0.0000) & (0.0000) & (0.0000) & (0.0000)  \\[3pt]
\multirow{2}{*}{400} 
& 0.0028 & 0.0227 & 0.0045 & 0.0021 & 0.0007 & 0.0006 & 1.0000 & 1.0000  \\
& (0.0001) & (0.0002) & (0.0001) & (0.0001) & (0.0000) & (0.0000) & (0.0000) & (0.0000)  \\[3pt]
\multirow{2}{*}{800} 
& 0.0016 & 0.0171 & 0.0037 & 0.0013 & 0.0004 & 0.0003 & 1.0000 & 1.0000  \\
& (0.0001) &(0.0000) & (0.0001) & (0.0001) & (0.0000) & (0.0000) & (0.0000) & (0.0000)  \\

\midrule

$T$ & \multicolumn{8}{c}{$N=200$, $K=3$, $R_D=2$, $R_S=3$} \\ [3pt]

\multirow{2}{*}{20} 
& 0.0053 & 0.0298 & 0.0059 & 0.0033 & 0.0013 & 0.0011 & 1.0000 & 1.0000  \\
& (0.0002) & (0.0003) & (0.0002) & (0.0002) & (0.0001) & (0.0000) & (0.0000) & (0.0000)  \\[3pt]
\multirow{2}{*}{25} 
& 0.0053 & 0.0272 & 0.0059 & 0.0032 & 0.0011 & 0.0009 & 1.0000 & 1.0000  \\
& (0.0002) & (0.0003) & (0.0001) & (0.0001) & (0.0000) & (0.0000) & (0.0000) & (0.0000)  \\[3pt]
\multirow{2}{*}{30} 
& 0.0053 & 0.0255 & 0.0059 & 0.0031 & 0.0010 & 0.0008 & 1.0000 & 1.0000  \\
& (0.0002) & (0.0003) & (0.0001) & (0.0001) & (0.0000) & (0.0000) & (0.0000) & (0.0000)  \\
\midrule

$K$ & \multicolumn{8}{c}{$N=400$, $T=25$, $R_D=2$, $R_S=3$} \\ [3pt]

\multirow{2}{*}{3} 
& 0.0028 & 0.0198 & 0.0045 & 0.0020 & 0.0005 & 0.0005 & 1.0000 & 1.0000  \\
& (0.0001) & (0.0001) & (0.0001) & (0.0001) & (0.0000) & (0.0000) & (0.0000) & (0.0000)  \\[3pt]
\multirow{2}{*}{4} 
& 0.0023 & 0.0198 & 0.0045 & 0.0020 & 0.0006 & 0.0005 & 1.0000 & 1.0000  \\
& (0.0001) & (0.0001) & (0.0001) & (0.0001) & (0.0001) & (0.0000) & (0.0000) & (0.0000)  \\[3pt]
\multirow{2}{*}{5} 
& 0.0019 & 0.0199 & 0.0045 & 0.0020 & 0.0007 & 0.0006 & 1.0000 & 1.0000  \\
& (0.0001) & (0.0002) & (0.0001) & (0.0001) & (0.0002) & (0.0001) & (0.0000) & (0.0000)  \\ 
\midrule

$R_S,R_D$ & \multicolumn{8}{c}{$N=200$, $T=20$, $K=3$} \\ [3pt]

\multirow{2}{*}{2,~3} 
& 0.0053 & 0.0298 & 0.0059 & 0.0033 & 0.0013 & 0.0011 & 1.0000 & 1.0000  \\
 & (0.0002) & (0.0003) & (0.0002) & (0.0002) & (0.0001) & (0.0000) & (0.0000) & (0.0000)  \\[3pt]
\multirow{2}{*}{4,~5} 
& 0.0068 & 0.0274 & 0.0157 & 0.0051 & 0.0021 & 0.0019 & 1.0000 & 1.0000  \\
& (0.0003) & (0.0003) & (0.0029) & (0.0004) & (0.0002) & (0.0001) & (0.0000) & (0.0000)  \\[3pt]
\multirow{2}{*}{6,~7} 
& 0.0082 & 0.0297 & 0.0343 & 0.0094 & 0.0034 & 0.0032 & 1.0000 & 1.0000  \\
& (0.0005) & (0.0003) & (0.0077) & (0.0010) & (0.0009) & (0.0011) & (0.0000) & (0.0000)  \\
\bottomrule
\end{tabular}
}
\label{table5}
\end{table}

\begin{table}[htbp]
    \centering
    \footnotesize
\setlength{\tabcolsep}{5pt}
    \captionsetup{skip=8pt}
    \caption{The True Positive Rate (TPR) and False Positive Rate (FPR) under different sparsity levels (standard deviations are in brackets).}
    \label{table9}
\resizebox{\textwidth}{!}{
    \begin{tabular}{ccccccccc}
        \toprule
        & \multicolumn{4}{c}{$s_0=0.3$} & \multicolumn{4}{c}{$s_0=0.5$} \\
        \cmidrule(lr){2-5} \cmidrule(lr){6-9}
        $N$ & \multicolumn{2}{c}{TPR} & \multicolumn{2}{c}{FPR} & \multicolumn{2}{c}{TPR} & \multicolumn{2}{c}{FPR} \\
        \cmidrule(lr){2-3} \cmidrule(lr){4-5}
        \cmidrule(lr){6-7}
        \cmidrule(lr){8-9}
        & \shortstack{Spa-Sim-\\STANE} & \shortstack{Spa-\\STANE} & \shortstack{Spa-Sim-\\STANE} & \shortstack{Spa-\\STANE} & \shortstack{Spa-Sim-\\STANE} & \shortstack{Spa-\\STANE} & \shortstack{Spa-Sim-\\STANE} & \shortstack{Spa-\\STANE}  \\
        \midrule
        
        \multirow{2}{*}{200} 
        & 0.8410 & 0.9360 & 0.0000 & 0.0000 & 0.8420 &  1.000 & 0.0000 & 0.0000  \\
        & (0.0157) & (0.0003) & (0.0000) & (0.0000) & (0.0129) & (0.0003) & (0.0000) & (0.0000)  \\ [3pt]
  
        \multirow{2}{*}{400} 
        & 0.8460 & 1.000 & 0.0000 & 0.0000 & 0.8520 & 1.000 & 0.0000 & 0.0000  \\
        & (0.0135) & (0.0000) & (0.0000) & (0.0000) & (0.0063) & (0.0000) & (0.0000) & (0.0000)  \\[3pt]

        \multirow{2}{*}{800} 
        & 0.8880 & 1.000 & 0.0000 & 0.0000 & 0.9010 & 1.000 & 0.0000 & 0.0000  \\
        & (0.0096) & (0.0000) & (0.0000) & (0.0000) & (0.0075) & (0.0000) & (0.0000) & (0.0000)  \\

        \bottomrule
    \end{tabular}
    }
    \label{table6}
\end{table}

\section{Empirical results}
\label{Empirical results}
\subsection{POLECAT data analysis}
In this section, we apply our proposed model to a real-world conflict relation dataset among different regions to demonstrate its practical utility in recovering both shared and time-specific latent positions, as well as the temporal group structure. Specifically, we analyze the POLECAT dataset of geopolitical interactions \cite{halterman2023plover}.
The POLECAT dataset records global political events and has been widely used in recent studies on dynamic networks (e.g., \cite{loyal2023eigenmodel}, \cite{baum2024doubly}). It totally contains 670,322 political interactions among 197 regions and has been recorded monthly from September 2022 to May 2024. These interactions are categorized into 16 types, which can be grouped into four broader categories: material conflict, verbal conflict, material cooperation, and verbal cooperation. In our analysis, we focus on conflict-related interactions between regions, as shifts in conflict dynamics are crucial for understanding the volatility of international political relations. Our goal is to assess whether the proposed method can effectively detect the time points at which political conflict patterns change, as captured by the temporal group structure of the dynamic components. 

We analyze conflict events over a 17-month period from January 2023 to May 2024, with a total of 56,023 edges observed. Among these 17 monthly networks, the first exhibits the highest density at 0.14, while the last has the lowest at 0.04. Moreover, the network density decreases year by year indicating that the conflict-related interactions among different regions are decreasing. Although the original dataset represents directed interactions—since political conflict is often initiated by one region toward another—we convert the directed networks into undirected ones by disregarding the directionality of the edges.

\begin{figure}[t]
    \centering
    \includegraphics[width=\textwidth]{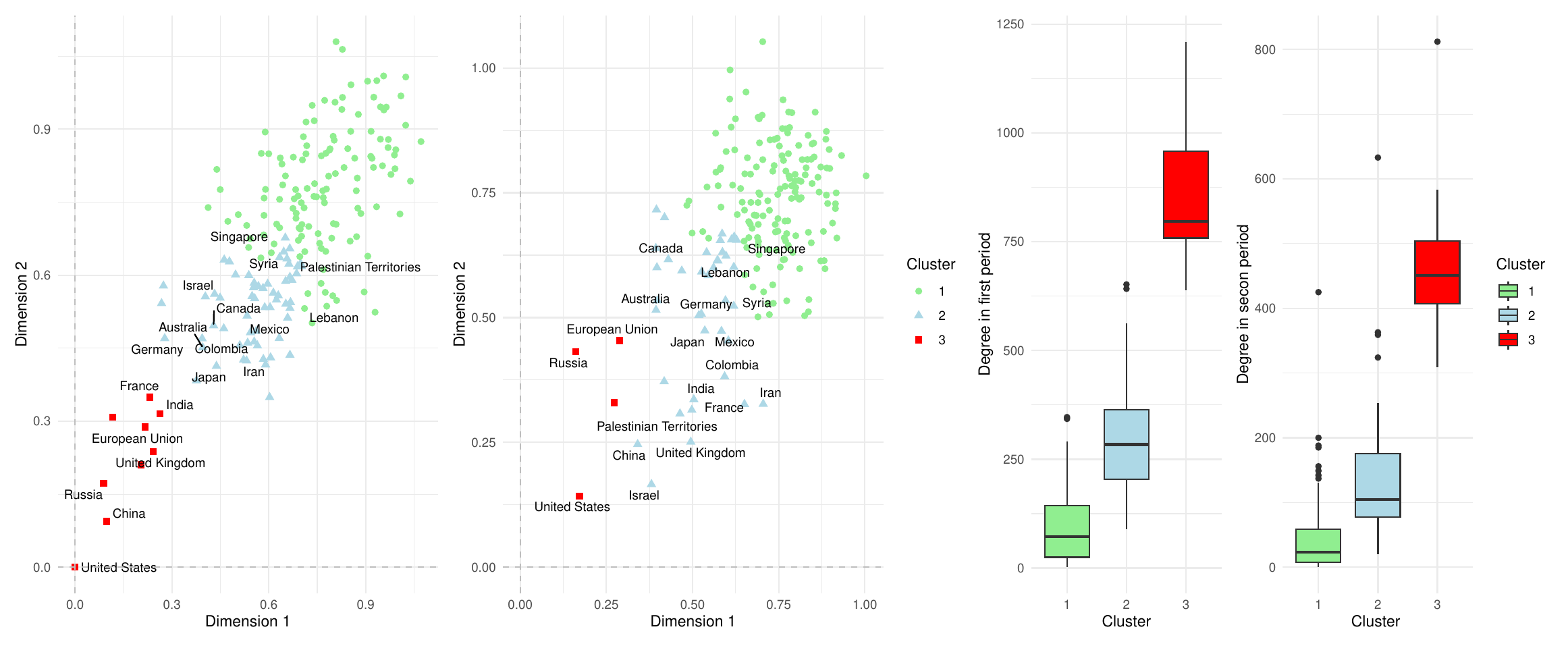}
    \caption{Estimated latent spaces for the POLECAT data: the first group and the second group are shown in the first two panels. The last two panels display boxplots of the average node degrees for each region across the two corresponding periods.}
    \label{Estimated U}
\end{figure}

We fit the proposed model to the POLECAT dataset and use BIC to determine the dimension of latent positions and the number of groups, resulting in $R_S=2$, $R_D=2$ and $K=2$. Our method partitions the entire timeline into 2 parts: January 2023 to September 2023, and October 2023 to May 2024. Interestingly, the identified change point—October 2023—coincides with a widely recognized escalation in geopolitical tensions in the Middle East, which received extensive global media attention. 
To gain further insights into the structure of international relationships, we apply the hierarchical clustering to the estimated latent position matrices $U^{(1)}$ and $U^{(2)}$. Figure \ref{Estimated U} presents the estimated latent positions for both periods, along with the average node degrees.
From the first two panels in Figure \ref{Estimated U}, we observe that both Israel and the Palestinian Territories exhibit substantial shifts in their latent positions between the two periods, indicating notable changes in their international interaction patterns. Since all estimated $\{V^{(t)}\}_{t=1}^T$ are negative, the inner product between Israel and western countries such as United States, Canada, and Germany increase from the first to the second period, suggesting growing alignment with Western nations. Meanwhile, the United States remains at the origin in the first period, highlighting the stability of its interaction patterns over time. The last two panels in Figure \ref{Estimated U} show the average degree by cluster. We find that regions in Cluster 3 exhibit the most intense conflict relations, followed by Cluster 2 and Cluster 1 in both periods.

\begin{table}[btp]
\captionsetup{skip=8pt}
\centering
\caption{Average link prediction results for all network layers on the POLECAT dataset (standard deviations are in brackets).}
\resizebox{\textwidth}{!}{
\begin{tabular}{c|cccccc}
\toprule
Metrics & MultiNeSS & M-GRAF & Sim-STANE & Spa-Sim-STANE & STANE & Spa-STANE \\
\midrule
\multirow{2}{*}{AuROC}  & 0.8376 & 0.8830 & 0.8963 & 0.8975 & 0.9192 & 0.9193 \\
& (0.0039) &  (0.0034) & (0.0048) & (0.0051) &  (0.0019) & (0.0019) \\
\multirow{2}{*}{AuPR} & 0.5009  & 0.5742 & 0.6001 & 0.6030 & 0.6297 & 0.6298\\& (0.0091) &  (0.0070) & (0.0075) & (0.0076) &  (0.0058) & (0.0058) \\
\multirow{2}{*}{MSE} & 0.0593  & 0.0526 & 0.0504 & 0.0503 & 0.0483 & 0.0482\\& (0.0010) &  (0.0006) & (0.0006) & (0.0006) &  (0.0005) & (0.0005) \\
\multirow{2}{*}{Log-Loss} & 0.2865  & 0.2158 & 0.1856 & 0.1853 & 0.1691 & 0.1691\\& (0.0074) &  (0.0042) & (0.0034) & (0.0035) &  (0.0002) & (0.0002) \\

\bottomrule
\end{tabular}
}
\label{aucs}
\end{table}

In practice, estimation of the latent positions is often an initial step, and the estimated model can be further used for downstream tasks on networks, for example, link prediction. Therefore, in order to demonstrate the performance of
the proposed method on link prediction,  we randomly remove 20\% entries of the adjacency matrices as missing and then use the remaining data to fit the model. For comparison, we also fit the methods we mentioned in section \ref{sim}. Then we predict link probabilities on those missing entries using the fitted parameters and node latent positions. 
The prediction result are evaluated under four metrics: Area under ROC curve (AuROC), Area under Precision-Recall curve (AuPR), Mean Square Error (MSE), and Log-Loss. The Log-Loss is defined as:
\(
-1/(T|\Phi|) \sum_{(i, j) \in \Phi} \sum_{t=1}^{T}\left\{A_{i j}^{(t)} \log \left(\widehat{P}_{i j}^{(t)}\right)+\left(1-A_{i j}^{(t)}\right) \log \left(1-\widehat{P}_{i j}^{(t)}\right)\right\},
\)
where $\Phi$ is the set of held-out edges (20\% randomly removed), and $\vert \Phi\vert$ denotes its cardinality. Table \ref{aucs} reports average results over 100 replications.
It is easy to find that the Spa-STANE outperforms all the other methods in terms of link prediction ability. Specifically, considering the group structure and sparsity of time specific latent positions has improve the accuracy of link prediction, which further supports the observed phenomenon that there exist time invariant shared structure of dynamic networks and it is critical to capture this network characteristics.

\subsection{Trade data analysis}
In this subsection, we apply our proposed model to the trade flow networks among different regions, namely the World Trade Web (WTW), also referred to as the International Trade Networks (ITN) \citep{gleditsch2002expanded}. Several studies in the existing literature have investigated the WTW dataset (e.g., \cite{cheung2020simultaneous}) in the context of dynamic network analysis. In brief, the dataset records trade flows between 190 regions from 1948 to 2000, consisting of the total annual imports and exports between each pair of regions. We construct dynamic networks by treating trade flows over one year periods as individual snapshots, and there is an edge between two regions if and only if both imports and exports between them are greater than zero during that year. Since we focus on undirected networks, we disregard the direction of imports and exports. The resulting dynamic networks consists of $T=53$ timepoints, and a total of 482,400 edges, with an average density of 0.25.

\begin{figure}[t]
    \centering
    \includegraphics[width=1\linewidth]{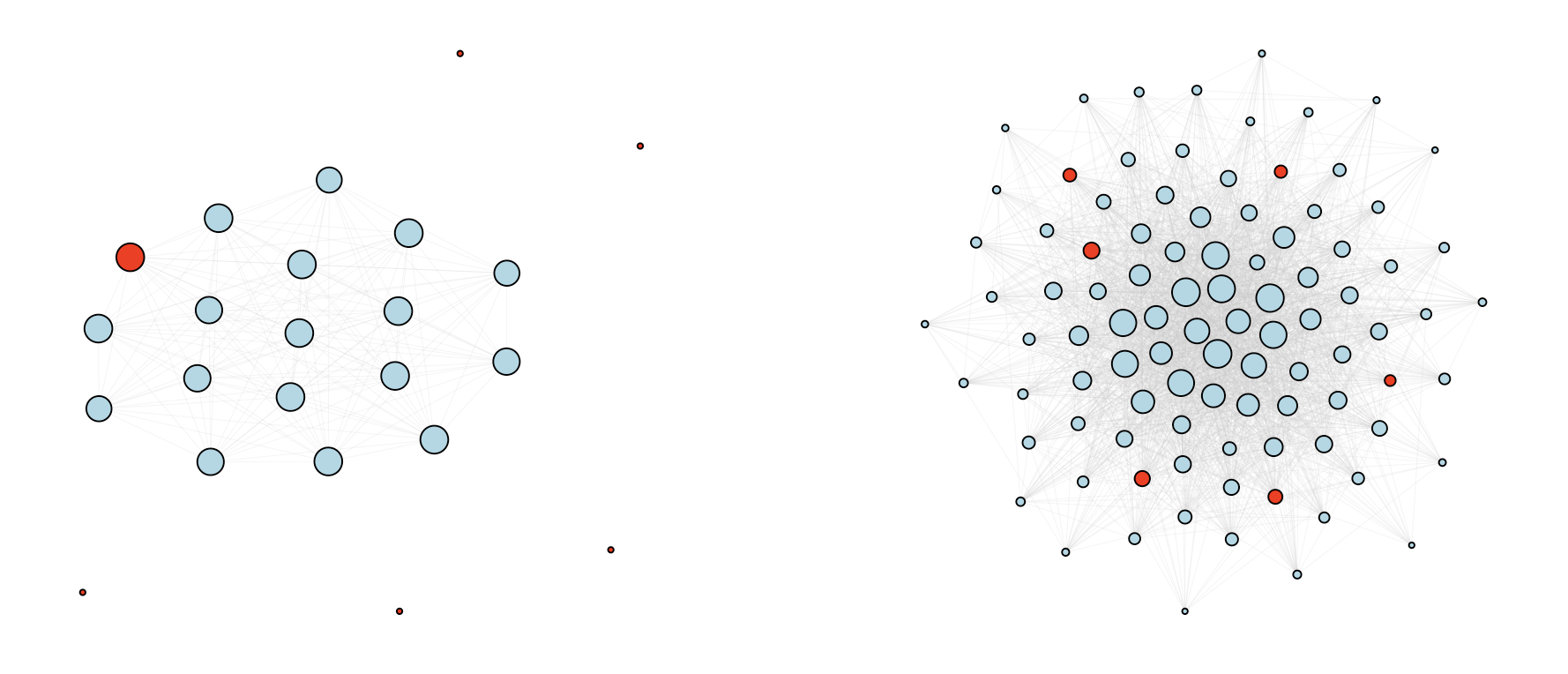 }
    \caption{The sub trade flow networks corresponding to the six African regions along with their neighbors 
and connecting edges during 1952(left) and 1967(right).}
    \label{fig:trade flow network}
\end{figure}

We use the proposed model to fit the trade flow  networks and use BIC criterion to determine the dimension of latent position and the number of groups, resulting in $R_S=3$, $R_D=2$ and $K=4$. Our method identifies the 3 change points and partitions the whole timeline into 4 parts, which is: 1948-1960, 1961-1974, 1975-1990, 1991-2000. The partition results are quite similar with the existing literature detecting the change point of trade flow networks \cite{cheung2020simultaneous}.
The first period 1948-1960 corresponds to the postwar reconstruction era, during which the establishment of General Agreement on Tariffs and Trade (GATT) in 1948 promoted tariff bindings and trade liberalization, gradually restoring trade flows among Western economies \citep{busch2003developing}. However, a substantial portion of Africa remained under colonial rule during this period and was therefore absent from the trade flow.
Between 1961 and 1974, trade liberalization accelerated, and an increasing number of African regions started to engage in global trade. To illustrate this, we select two time points, one from the first period and one from the second period, then plot the corresponding networks for the six African regions along with their neighbors and connecting edges. Figure \ref{fig:trade flow network} presents these two networks, where the red circles denote the African regions, and the size of each node reflects its degree. It can be observed that from the first to the second period, the trade flow intensity among the African regions increases gradually. Moreover, through the Kennedy Round of GATT, which led to substantial tariff reductions and deeper regional integration in Europe \citep{coppolaro2011us}.
From 1975 to 1990, trade rules expanded during the Tokyo Round, while regional trade agreements became more prominent and global value chains began to emerge \citep{coppolaro2018shadow}. Finally, the years 1991–2000 reflect the worldwide economic globalization, and the establishment of the World Trade Organization (WTO) in 1995 institutionalized a broader set of rules covering services and intellectual property, further densifying and globalizing the trade network \citep{deardorff2002you}. The partition of the entire timeline based on our proposed model  gives the meaningful  interpretation of the evaluation of the global trade. Besides the interesting classification results, our Spa-STANE framework also reveals that the group-specific latent embedding vectors of some regions shrink to zero in the last period, indicating that their connection patterns with other regions remain stable during 1991–2000. These regions include, but are not limited to, the United States, France, the Netherlands, and Germany, which are countries with well-developed economies and active trade connections.

We also evaluate the link prediction accuracy of the proposed method on world trade flow networks. Similar with the previous example, 20\% of the entries in the adjacency matrices are randomly masked to simulate missing links, and the model is fitted using the remaining observations. The prediction performance is summarized in Table \ref{aucs for ex2}. It can be observed that the proposed STANE framework achieves best link prediction performance in terms of AuROC and Log-Loss compared with other methods, demonstrating its adaptability and effectiveness in modeling trade flow networks. It is worth noting that the prediction performance of M-GRAF is comparable to that of the STANE framework. However, it does not account for the group structure or the sparsity of the time-varying components, resulting in less interpretability when modeling real-world dynamic networks. 

\begin{table}[btp]
\captionsetup{skip=8pt}
\centering
\caption{Average link prediction results for all network layers on the trade flow data (standard deviations are in brackets).}
\resizebox{\textwidth}{!}{
\begin{tabular}{c|cccccc}
\toprule
Metrics & MultiNeSS & M-GRAF & Sim-STANE & Spa-Sim-STANE & STANE & Spa-STANE \\
\midrule
\multirow{2}{*}{AuROC}  & 0.9134 & 0.9465 & 0.9325 & 0.9310 & 0.9481 & 0.9482 \\
& (0.0011) &  (0.0010) & (0.0010) & (0.0008) &  (0.0010) & (0.0010) \\
\multirow{2}{*}{AuPR} & 0.8594  & 0.8932 & 0.8467 & 0.8441 & 0.8778 & 0.8783\\& (0.0019) &  (0.0015) & (0.0013) & (0.0016) &  (0.0019) & (0.0021) \\
\multirow{2}{*}{MSE} & 0.0835  & 0.0768 & 0.0891 & 0.0906 & 0.0793 & 0.0792\\& (0.0008) &  (0.0005) & (0.0004) & (0.0004) &  (0.0007) & (0.0007) \\
\multirow{2}{*}{Log-Loss} & 0.4378 & 0.2976  & 0.2971 & 0.3039 & 0.2617 & 0.2611\\& (0.0072) &  (0.0055) & (0.0072) & (0.0021) &  (0.0022) & (0.0023) \\

\bottomrule
\end{tabular}
}
\label{aucs for ex2}
\end{table}


\section{Conclusions}
We introduce STANE, a novel joint network embedding framework for dynamic networks that simultaneously captures shared and time-specific structures with temporal clustering. To further enhance its ability to model localized changes over time, we propose Sparse STANE, an extension that explicitly incorporates sparse perturbation modeling. Extensive experiments on synthetic and real-world datasets highlight the strengths of the STANE framework in recovering temporal structures, detecting structural shifts, and providing insightful representations of dynamic networks.

There are several directions for future research. First, extending STANE to handle continuous-time dynamic networks, where interactions occur at irregular and potentially high-resolution time points rather than in discrete snapshots, would broaden its applicability. Second, incorporating node or edge covariates into the embedding framework could enhance interpretability and prediction. Finally, developing scalable versions of the algorithm for massive networks, possibly through distributed computation, is a promising avenue.

\section{Disclosure statement}\label{disclosure-statement}

The authors have no conflicts of interest exist.

\section{Data Availability Statement}\label{data-availability-statement}

Deidentified data have been made available at the following URL: \url{https://github.com/baihairi/Shared-and-Time-specific-Adaptive-Network-Embedding}.

\phantomsection\label{supplementary-material}
\bigskip

\begin{center}

{\large\bf SUPPLEMENTARY MATERIAL}

\end{center}

\begin{description}
\item[Title:]
Appendix to ``Common-Individual Embedding
for Dynamic Networks with Temporal Group
Structure''. (.pdf type)

The supplementary material consists of three sections. Section A provides the proofs of Proposition 1 and the theoretical results presented in Section 3. Section B describes the initialization procedure of Algorithm 1 in detail. Section C presents additional simulation results.

\end{description}

\bibliography{main}

\end{document}